\documentclass[journal]{IEEEtran}
\IEEEoverridecommandlockouts
\usepackage{cite}
\usepackage{graphicx}
\usepackage{psfrag}
\usepackage{subfigure}
\usepackage{url}
\usepackage{stfloats}
\usepackage{amsfonts,amssymb,amsmath,bm,paralist,theorem,cite,ifthen,color}
\usepackage{array}
\usepackage{caption}
\usepackage{calc}
\usepackage{upgreek}
\usepackage{subfig}
\usepackage{longtable}
\usepackage{stfloats}
\usepackage{graphicx}
\usepackage{algorithm,algpseudocode}

\newtheorem{lemma}{Lemma}

\begin{document}

%\title{Cognitive Coexistence of Ad-hoc and Cooperative Relay Networks}%: Resource Allocation for Interference Mitigation}
\title{Spectrum Sharing between Cooperative Relay and Ad-hoc Networks: Dynamic Transmissions under Computation and Signaling Limitations}
%\title{Spectrum Sharing between Ad-hoc and Cooperative Relay Networks: Interference Mitigation, Resource Allocation and Real-time Implementations}
%\title{Spectrum Sharing between Ad-hoc and Cooperative Relay
%Networks: Interference Prediction, Resource Allocation and Real-time
%Implementations}

\author{\authorblockN{Yin~Sun,~Xiaofeng Zhong,~Yunzhou Li,~Shidong~Zhou and~Xibin
Xu\\}
\authorblockA{State Key Laboratory on Microwave and Digital
Communications\\
Tsinghua National Laboratory for Information Science and
Technology\\
Department of Electronic Engineering, Tsinghua University, Beijing,
100084, China.\\
E-mail: sunyin02@gmail.com,
\{zhongxf,liyunzhou,zhousd,xuxb\}@tsinghua.edu.cn.}

\thanks{This work was supported by National S\&T Major Project (2008ZX03O03-004), National Basic Research Program of China
(2007CB310608), China's 863 Project (2009AA011501), National Natural
Science Foundation of China (60832008) and Tsinghua-Qualcomm Joint
Research Program.} }

\maketitle

\begin{abstract}
This paper studies a spectrum sharing scenario between a cooperative relay network (CRN) and a nearby ad-hoc network. In particular, we consider a dynamic spectrum access and resource allocation problem of the CRN. Based on sensing and predicting the ad-hoc transmission behaviors, the ergodic traffic collision time between the CRN and ad-hoc network is minimized subject to an ergodic uplink throughput requirement for the CRN. %The formulated design problem is a difficult nonconvex optimization problem with no closed-from expression for the objective function. By carefully analyzing the problem structure, we show how this problem can be reformulated as a convex problem. A low-complexity Lagrangian optimization method is then proposed to solve the considered design problem efficiently.

We focus on real-time implementation of spectrum sharing policy under practical computation and signaling limitations. In our spectrum sharing policy, most computation tasks are accomplished off-line. Hence, little real-time calculation is required which fits the requirement of practical applications. Moreover, the signaling procedure and computation process are designed carefully to reduce the time delay between spectrum sensing and data transmission, which is crucial for enhancing the accuracy of traffic prediction and improving the performance of interference mitigation. %The key idea to achieve these benefits is making use of the analytical structure of the optimal solution.
The benefits of spectrum sensing and cooperative relay techniques are demonstrated by our numerical experiments.
\end{abstract}

\IEEEpeerreviewmaketitle

\section{Introduction}
In recent years, spectrum sharing between heterogeneous wireless
networks has been studied intensively as a crucial technology for improving
network spectrum efficiency
\cite{Zhao_survey} and network
capacity
\cite{Chandrasekhar_Femto_survey08}. %,Servey_spectrum_utilization09,Hybrid_Cellular_AdHoc_Ton10
Traffic prediction based spectrum access
polices were proposed in \cite{Geirhofer_ComMag,Geirhofer_Jsac08,Zhaoqianchuan_TSP08,GeirhoferExperiment09,You_Xu10,
LifengLai_MAC11,ZhaoTong_Jsac11}, %,,Zhao_Jsac07, Xi_ZhangJsac08,
where the cognitive transmitter detects and predicts the primary user's (PU) transmission behaviors and
transmits signals opportunistically to avoid
collisions with the PU's traffic. %While these works focus on MAC-layer spectrum access,
%there are works focusing on physical-layer resource allocation of the
%secondary systems, under constraints that limit the induced
%interference to the primary receivers; see
%\cite{Peng_Wang07,RuiZhangJSTSP08,PowerOFDMCogTSP09} and also
%\cite{RuiZhangMag10} for an overview.
Joint optimization of spectrum access and resource allocation based on traffic prediction has been proposed in \cite{Geirhofer_Mobile_computing} for an open sharing model \cite{Zhao_survey} that considers spectrum sharing between an uplink system and an ad-hoc network. In \cite{YinSunICC2010}, cooperative relay technique was utilized to improve the spectrum sharing performance.

%The rapid development of wireless communications has made frequency
%spectrum a very scare resource. Recently, spectrum sharing
%techniques based on interference prediction have received much
%attention. Cognitive medium access strategies were analyzed in
%\cite{Zhao2}-\cite{Geirhofer3}, which did not consider the power
%allocation of the secondary users. Interference prediction based
%resource allocation was studied for cognitive OFDMA system
%\cite{Geirhofer1}.

However, some implementation issues were rarely considered in these
studies. First, determining the resource allocation policy in real-time
can be computationally quite demanding for realistic wireless communication systems \cite{cellular_adhoc_tradeoffjsac09}. Second, spectrum sensing and channel estimation are usually performed at spatially separate nodes, which requires to exchange their obtained information before solve the resource allocation problem. The resultant signaling procedure and the computation of resource allocation solution would cause a large time delay between spectrum sensing and data transmission, which would degrade the accuracy of traffic prediction and cause unexpected traffic collisions between the networks operating in the same spectrum. Therefore, resource allocation policies with little real-time calculation and small sensing-transmission delay are of great interest for practical applications.

In this paper, we study spectrum sharing between a cooperative relay
network (CRN) and an ad-hoc network, as illustrated in Fig.
\ref{fig1}. The relay assists the transmissions from
the mobile terminal (MT) to the base station (BS) to achieve higher uplink throughput. In order to communicate with the distant BS, the MT and relay would transmit
signals with peak powers, which induce strong
interference to nearby ad-hoc links. The ad-hoc transmitters (e.g., wireless sensor nodes) have relative low transmission powers due to their short
communication ranges, and thus their interference to the relay and
BS can be treated as noise. Such an asymmetrical interference
scenario is known as the ``near-far effect"
\cite{Chandrasekhar_Femto_survey08}.

We consider a joint spectrum access and resource allocation problem of the CRN, where the ergodic traffic collision time between the CRN and ad-hoc network is minimized subject to an ergodic uplink throughput constraint for the CRN. The formulated design problem is a difficult nonconvex optimization problem with no closed-form
expression for the objective function. By carefully analyzing the problem structure, we show how this problem can be reformulated as a convex problem. A low-complexity Lagrangian optimization method is used to solve the considered design problem efficiently. Then, a real-time implementation policy is proposed which requires little real-time calculation and has small sensing-transmission delay. Finally, numerical results are provided to show the benefits of our spectrum sharing policy.

\section{System model}\label{sec13}
\begin{figure}[!t] \centering
    \resizebox{0.45\textwidth}{!}{
    \includegraphics{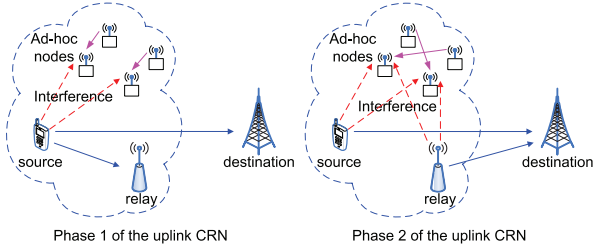}}
    \caption{System setup of the spectrum sharing between cooperative relay and
     ad-hoc networks.}
    \label{fig1}
    \vspace{-0.3cm}
\end{figure}
%The system model is illustrated in Fig. \ref{fig1}(a), where an
%uplink CRN with high transmission power generates strong
%interference to nearby victim ad-hoc receivers. %Each sub-channel can be a
%frequency band with certain bandwidth, or a group of continuous
%sub-carriers (a sub-band) of an orthogonal frequency division
%multiplexing (OFDM) system \cite{Zhao1}.
The CRN operates in frames with duration $T_f$. Each frame comprises
$N$ sub-channels in frequency domain, denoted by the set $\mathcal
{N}=\{1,2,\cdots,N\}$. We assume that the wireless channels of source-relay (S-R), source-destination (S-D), and relay-destination (R-D) links are block-faded, which vary across the frames in a stationary and ergodic manner. The channel gain normalized by the interference plus noise power of these links
are denoted by $g_n^{s,r}, g_n^{r,d},g_n^{s,d}$,
respectively, for the $n$-th sub-channel. %We assume that the channel state is perfectly estimated at the corresponding receivers, and the destination (base station, BS) acquires the channel quality information (CQI) through
%prediction before data transmission, if the wireless channel varies slowly across the frames \cite{ZhangYan_prediction}.

In practice, the relay node operates in a half-duplex mode. Therefore, each frame consists of 2 phases: In Phase 1, the source transmits signal to the relay and destination via a broadcast channel; in Phase 2, the source transmits a new information message, and, at the same time, the relay uses the DF relaying strategy to forward its received information message in Phase 1 to the destination, which forms a multiple-access channel. These operations are illustrated in Fig. \ref{fig1}. The time durations of Phase 1 and Phase 2 are set to $\alpha T_f$ and $(1-\alpha)T_f$, respectively, where $\alpha\in(0,1)$.

The ad-hoc links operate in $M$ non-overlapping frequency bands
denoted by the set $\mathcal {M}=\{1,2,\cdots,M\}$ and the $m$-th
ad-hoc band overlaps with a set of sub-channels given by $\mathcal
{N}_m$ ($\mathcal {N} = \bigcup_{m = 1}^M\mathcal {N}_m$ and
$\mathcal {N}_m\bigcap\mathcal {N}_l = {\O}$ if $m\neq l$).
The ad-hoc traffic in the $m$-th band is modeled by a strictly
stationary, ergodic and independent binary continuous-time Markov chain (CTMC)
$X_m(t)$, where $X_m(t)=1$ ($X_m(t)=0$) represents an ACTIVE (IDLE) state at time $t$. The holding (or sojourn) periods of ACTIVE and IDLE states are exponentially
distributed with rate parameters $\lambda$ and $\mu$, respectively. The probability
transition matrix of the CTMC model of Band $m$ is given by \cite[p.
391]{BK:Resnick}
\begin{equation} \label{eq15}
P(t)\!=\!\frac{1}{\lambda\!+\!\mu}
\!\!\left[\!\!\begin{array}{l l}\!\mu\!+\!\lambda
e^{-(\lambda\!+\!\mu)t}\!\!&\!\!\lambda\!-\!\lambda
e^{-(\lambda\!+\!\mu)t}\!\\\!\mu\!-\!\mu e^{-(\lambda\!+\!\mu)t}\!\!&\!\!\lambda\!+\!\mu
e^{-(\lambda\!+\!\mu)t}\!\end{array}\!\!\!\right]\!\!,
\end{equation} where the element in the $(i+1)$-th row and $(j+1)$-th column of $P(t)$ stands for the transition probability $\Pr\{X_m(t+\tau)=j|X_m(\tau)=i\}$ for $i, j\in\{0,1\}$.
This CTMC model has been considered in many spectrum sharing studies
including theoretical analysis and hardware tests; see
\cite{Geirhofer_ComMag,Zhaoqianchuan_TSP08,Geirhofer_Jsac08,You_Xu10,ZhaoTong_Jsac11,Geirhofer_Mobile_computing,LifengLai_MAC11,GeirhoferExperiment09,YinSunICC2010}.
%In practice, the parameters $\lambda$ and $\mu$ can be estimated
%by monitoring the ad-hoc traffic in idle frames
%\cite{Geirhofer_ComMag}.

The source and relay detect the ACTIVE/IDLE state of each
ad-hoc band at the start of both Phase 1 and Phase 2. The sensing outcome of the two phases are denoted by $X_m(0)=x_m\in\{0,1\}$ and $X_m(\alpha T_f)=y_m\in\{0,1\}$, respectively. Perfect sensing and negligible sensing overhead are assumed in this
paper.

\section{Problem formulation}
Let us define
$\bm \omega\triangleq\{g_n^{s,r},g_n^{r,d},g_n^{s,d},x_m,y_m,n\in\mathcal
{N},m\in\mathcal {M}\}$ as the network state information (NSI). The dynamic transmission parameters are determined by the instant NSI $\bm \omega$.
Suppose that the source and relay nodes can switch on and off their transmissions freely over each sub-channel, and may transmit only in part of the time during Phase 1 and Phase 2. Let $\mathbb{I}_n^{(1)}(\bm \omega)\subseteq[0,\alpha T_f]$ denote the set of transmission time of the source over Sub-channel $n$ in Phase 1, and $\mathbb{I}_n^{(2)}(\bm \omega)\subseteq[\alpha T_f,T_f]$ denote that of the source and relay in Phase 2, for $n=1,\ldots,N$. $\mathbb{I}_n^{(1)}(\bm \omega)$ and $\mathbb{I}_n^{(2)}(\bm \omega)$ each may be \emph{a union of several disjoint transmission time intervals}. %Then, $\mathbb{I}_n^{(1)}(\bm \omega)$ and $\mathbb{I}_n^{(2)}(\bm \omega)$ determine the spectrum access polices of the CRN in Phase 1 and Phase 2, respectively.
We utilize the words ``traffic collision'' to represent the event that both the CRN and ad-hoc network are transmitting in the same spectrum band at the same time. In \cite{YinSunTSP11}, we showed that ergodic traffic collision time between the two networks is given as
\begin{eqnarray}\label{eq105}
\!\!\!\!\!\!\!\!\!\!\!\!\!&&\overline{I}
\!=\!\mathbb{E}_{\bm\omega}\!\left\{\sum_{m=1}^M\!\!\left[\!\int_{\bigcup_{n\in\mathcal
{N}_m}\mathbb{I}_n^{(1)}(\bm \omega)}\!\!\!\!
\Pr\!\left\{X_m(\sigma)\!=\!1|X_m(0)\!=\!x_m\right\} d\sigma\!\!\!\!
\right.\right.\nonumber\\
\!\!\!\!\!\!\!\!\!\!\!\!\!&&\left.\left.+
\int_{\bigcup_{n\in\mathcal {N}_m}\mathbb{I}_n^{(2)}(\bm \omega)}
\!\!\!\!\Pr\left\{X_m(\sigma)=1|X_m(\alpha T_f)=y_m\right\}
d\sigma\right]\right\}\!,\!
\end{eqnarray}
which is proportional to the transmission error probability of the ad-hoc network in strong interference scenarios \cite{Geirhofer_Jsac08}.

Let $\pi(S)$ represents the size (measure) of set $S$; for example, $\pi([a,b])=b-a$. Thus, the transmission time fractions of the CRN are determined as $\theta_n^{(1)}(\bm \omega)={\pi(\mathbb{I}_n^{(1)}(\bm \omega))}/{T_f}$ and $\theta_n^{(2)}(\bm \omega)={\pi(\mathbb{I}_n^{(2)}/(\bm \omega))}{T_f}$, respectively, for Phase 1 and Phase 2 of the frame.
Then, the ergodic achievable rate of the CRN can be expressed as \cite{YinSunTSP11}
\begin{eqnarray}\label{eq47}
\!\!\!\!\!\!\!\!\!\!\!\!\!\!\!\!\!\!\!\!\!&&\overline{R}_{DF}
\nonumber\\
\!\!\!\!\!\!\!\!\!\!&&\!=\!W\!\min\!\sum_{n=1}^N\!\mathbb{E}_{\bm \omega}\left[{\theta_n^{(1)}}(\bm \omega)
\log_2\!\!\left(\!1\!+\!\frac{
P_{s,n}^{(1)}(\bm \omega)\max\{g_n^{s,r},g_n^{s,d}\}}
{\theta_n^{(1)}(\bm \omega)}\!\right)\right.\!\!\!\nonumber\\
\!\!\!\!\!\!\!\!\!\!&&~~\left.+{\theta_n^{(2)}}(\bm \omega)
\log_2\!\left(\!1\!+\!\frac{
P_{s,n}^{(2)}(\bm \omega)g_n^{s,d}}{\theta_n^{(2)}(\bm \omega)}\right)\!\right]\!,\!\!\nonumber\\
\!\!\!\!\!\!\!\!\!\!&&~~~\sum_{n=1}^N\mathbb{E}_{\bm \omega}\left[
{\theta_n^{(1)}(\bm \omega)} \log_2\left(1+\frac{
P_{s,n}^{(1)}(\bm \omega)g_n^{s,d}}{\theta_n^{(1)}(\bm \omega)}\right)\right.\nonumber\\
\!\!\!\!\!\!\!\!\!\!&&~~\left.\left.+
\theta_n^{(2)}(\bm \omega)
\log_2\left(1+\frac{P_{s,n}^{(2)}(\bm \omega)g_n^{s,d}+
P_{r,n}(\bm \omega)g_n^{r,d}}{\theta_n^{(2)}(\bm \omega)}\right)\right]\right\}\!.
\end{eqnarray} Note that this ergodic rate can be achieved in slow-fading environment by means of queuing at the relay node. Moreover, it is a concave function of $\{P_{s,n}^{(1)}(\bm \omega),P_{s,n}^{(2)}(\bm \omega),P_{r,n}(\bm \omega),\theta_n^{(1)}(\bm \omega),$ $\theta_n^{(2)}(\bm \omega),n\in\mathcal {N}\}$, since the perspective of a concave function is also concave \cite[p. 89]{BK:BoydV04}.

The joint spectrum access and resource allocation problem the CRN is formulated as
\begin{eqnarray}
(\sf P)~\min_{\substack{P_{s,n}^{(1)}(\bm \omega),P_{s,n}^{(2)}(\bm \omega),P_{r,n}(\bm \omega),\mathbb{I}_n^{(1)}(\bm \omega),\mathbb{I}_n^{(2)}(\bm \omega),\\\theta_n^{(1)}(\bm \omega),\theta_n^{(2)}(\bm \omega),~n=1,\ldots,N}}
&~\overline{I}
\end{eqnarray}
\begin{figure}[!t]
    \centering
        \resizebox{0.45\textwidth}{!}{\includegraphics{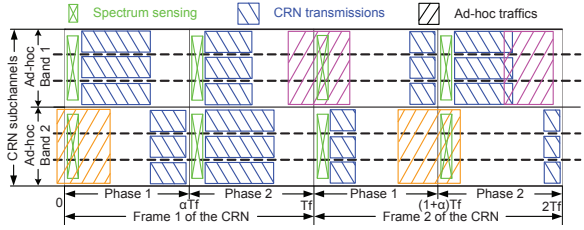}}
        \caption{Time-frequency transmission structure by Lemma \ref{lem3}. }%Here the sensing outcomes of frame 1 are $x_1=0$, $x_2=1$, $y_1=0$ and $y_2=0$, and the the sensing outcomes of frame 2 are $x_1=1$, $x_2=0$, $y_1=0$ and $y_2=1$.}
        \label{fig4}
        \vspace{-0.3cm}
\end{figure}
\vspace{-0.5cm}
\begin{eqnarray}
\!\!\!\!\!\!\!\!\!\!\!\!&&\!\!\!\!\!\!\!\!\!\!\text{s.t.~} \overline{R}_{DF}\geq R_{\min}\label{eq111}\\
\!\!\!\!\!\!\!\!\!\!\!\!&&\mathbb{E}_{\bm \omega}\left\{\sum_{n = 1}^N \left[P_{s,n}^{(1)}(\bm \omega)+ P_{s,n}^{(2)}(\bm \omega)\right]\right\}\leq P^s_{\max}\label{eq100}\\
\!\!\!\!\!\!\!\!\!\!\!\!&&\mathbb{E}_{\bm \omega}\left\{\sum_{n = 1}^N P_{r,n}(\bm \omega)\right\}\leq P^r_{\max}
\end{eqnarray}
\begin{eqnarray}
\!\!\!\!\!\!\!\!\!\!\!\!&&P_{s,n}^{(1)}(\bm \omega),P_{s,n}^{(2)}(\bm \omega),P_{r,n}(\bm \omega)\geq 0\label{eq4}\\
\!\!\!\!\!\!\!\!\!\!\!\!&&\mathbb{I}_n^{(1)}(\bm \omega)\subseteq[0,\alpha T_f],~~\mathbb{I}_n^{(2)}(\bm \omega)\subseteq[\alpha T_f,T_f]\label{eq44}\\
\!\!\!\!\!\!\!\!\!\!\!\!&&\pi(\mathbb{I}_n^{(1)}(\bm \omega)) = \theta_n^{(1)}(\bm \omega)T_f,~~\pi(\mathbb{I}_n^{(2)}(\bm \omega))=\theta_n^{(2)}(\bm \omega)T_f. \label{eq64}
\end{eqnarray}
%\begin{eqnarray} \!\!\!\!\!\!\!\!\!(\sf P)~\textrm{arg}
%\min\limits_{P_{s,n}^{(1)}(\bm \omega),P_{s,n}^{(2)}(\bm \omega),P_{r,n}(\bm \omega),\mathbb{I}_n^{(1)}(\bm \omega),\mathbb{I}_n^{(2)}(\bm \omega)}\!\!\!\!
%&&\!\!\!\! I_{average}
%\end{eqnarray}
%subject to (\ref{eq108})-(\ref{eq115}) and the average rate
%constraint
%\begin{eqnarray}\label{eq111}
%R_{DF,average}\geq R_{\min}.
%\end{eqnarray}
%It is hard to solve the problem $(P1)$, because it is a nonconvex
%optimization problem and the interference metric (\ref{eq105})
%involves integrations which do not have closed-form expressions.
\section{The solution to the problem $(P)$}
Problem $(\sf P)$ is difficult to solve mainly because it is hard to determine the sets $\mathbb{I}_n^{(1)}(\bm \omega)$ and
$\mathbb{I}_n^{(2)}(\bm \omega)$ and thus the objective function $\overline{I}$ has no closed-form expression in general. %In the next section, we will show how these issues can be resolved and eventually $(\sf P)$ can be reformulated as a convex optimization problem.
Fortunately, these issues can be resolved and eventually $(\sf P)$
can be reformulated as a convex optimization problem, as we
present in the following.
\subsection{Transformation of $(P)$ to a convex problem}
In \cite{YinSunTSP11}, we show that the optimal spectrum access should satisfy the following two principles:
\begin{enumerate}
\item The source and relay nodes should transmit as soon (late)
as possible if the sensing outcome is IDLE (ACTIVE);
\item The CRN should have identical spectrum access policy for all sub-channels in $\mathcal {N}_m$; that is, $\mathbb{I}_p^{(i)}=\mathbb{I}_q^{(i)}$ for all $p,q\in\mathcal {N}_m$ and $i\in\{1,2\}$.
\end{enumerate}
Let us define
\begin{eqnarray}
\theta_m^{(i)}(\bm\omega)\triangleq\max\left\{\theta_n^{(i)}(\bm\omega),n\in \mathcal {N}_m\right\},
\end{eqnarray}
for $m=1,\ldots,M$ and $i=1,2$. The above principles are formalized in the following Lemma:

\begin{lemma} \cite{YinSunTSP11}\label{lem3}
For any given transmission time fractions
$\{\theta_m^{(1)}(\bm\omega)\in[0,\alpha],\theta_m^{(2)}(\bm\omega)\in[0,1-\alpha]\}_{m=1}^M$,
we have that:
\begin{enumerate}
\item The
optimal spectrum access policy of Phase 1 is given by $\mathbb{I}_n^{(1)}(\bm\omega)=[0,\theta_m^{(1)}(\bm\omega)T_f]$ $( \mathbb{I}_n^{(1)}(\bm\omega)=[(\alpha-\theta_m^{(1)}(\bm\omega))T_f, \alpha T_f])$ for all $n\in\mathcal {N}_m$, if the
sensing outcome of Phase 1 is $x_m = 0$ $(x_m = 1)$;
\item The
optimal spectrum access policy of Phase 2 is given by $\mathbb{I}_n^{(2)}(\bm\omega)=[\alpha
T_f,(\alpha+\theta_m^{(2)}(\bm\omega))T_f]$
$(\mathbb{I}_n^{(2)}(\bm\omega)=[(1-\theta_m^{(2)}(\bm\omega))T_f, T_f])$ for all $n\in\mathcal {N}_m$, if the
sensing outcome of Phase 2 is $y_m = 0$ $(y_m = 1)$,
\end{enumerate}
\end{lemma}An example of the spectrum access policy in Lemma \ref{lem3} is shown in Fig. \ref{fig4}. According to Lemma \ref{lem3}, each
term inside the expectation in \eqref{eq105} can be greatly simplified. For $\theta\in[0,\alpha]$, define the functions
\begin{eqnarray}\label{eq16}
\!\!\!\!\!\!\!\!\!\!\!\phi_{(1)}
(\theta;0)\!\!\!\!\!\!\!\!\!\!\!&&=\!\int_{[0,\theta T_f]} \Pr(X_m(t)\!=\!1|X_m(0)=0)d
t\!\nonumber\\
&&\!\!\!\!\!\!\!\!\!\!\!\!\!\!\!\!\!\!\!\!\!\!\!\!=\!\frac{\lambda T_f}{\lambda+\mu}\left\{\theta +\frac{1}{(\lambda+\mu)T_f}\!\!\left[\!e^{-({\lambda+\mu})\theta T_f}-1\!\right]\!\!\right\},
\end{eqnarray}
\begin{eqnarray}
\!\!\!\!\!\phi_{(1)}(\theta;1)\!\!\!\!\!\!\!\!\!\!\!&&=
\int_{[(\alpha-\theta) T_f,\alpha T_f]}\Pr (X_m(t)=1|X_m(0)=1)d t\nonumber\\
&&\!\!\!\!\!\!\!\!\!\!\!\!\!\!\!\!\!\!\!\!\!\!\!\!=\frac{\lambda T_f}{\lambda+\mu}\left\{\theta
+\frac{{\mu }/{\lambda}}{(\lambda+\mu)
T_f}e^{-({\lambda+\mu})\alpha
T_f}\left[e^{({\lambda+\mu})\theta T_f}\!-\!1\!\right]\!\!\right\}\!,
\label{eq166}
\end{eqnarray}
and for $\theta \in[0, \alpha]$, define the functions
\begin{eqnarray}\label{eq78}
\!\!\!\!\!\!\!\!{\phi}_{(2)}(\theta;0)\!\!\!\!\!\!\!\!\!\!\!&&=\!\int_{[\alpha
T_f,(\theta+\alpha) T_f]} \Pr(X_m(t)\!=\!1|X_m(\alpha T_f)=0)d
t\!\nonumber\\
&&\!\!\!\!\!\!\!\!\!\!\!\!\!\!\!\!\!\!\!\!\!\!\!\!=\!\frac{\lambda T_f}{\lambda+\mu}\left\{\theta +\frac{1}{(\lambda+\mu)T_f}\!\!\left[\!e^{-({\lambda+\mu})\theta T_f}-1\!\right]\!\!\right\},
\\
\!\!\!\!\!\!{\phi}_{(2)}(\theta;1)\!\!\!\!\!\!\!\!\!\!\!&&=\int_{[T_f-\theta T_f,T_f]}
\Pr(X_m(t)=1|X_m(\alpha T_f)=1)d
t\!\nonumber\\
&&\!\!\!\!\!\!\!\!\!\!\!\!\!\!\!\!\!\!\!\!\!\!\!\!=\!\frac{\lambda T_f}{\lambda\!+\!\mu}\left\{\theta +\!\frac{{\mu}/{\lambda}}{(\lambda\!+\!\mu)T_f}e^{-({\lambda+\mu})(1-\alpha)T_f}\!\!\left[e^{({\lambda+\mu})\theta T_f}\!-\!1\!\right]\!\!\right\}\!.\label{eq79}
\end{eqnarray}
It is easy to prove that the functions ${\phi}_{(i)}(\theta;x)$ are strictly convex in $\theta$ by considering their secondary derivations.
Then, the interference metric in \eqref{eq105} can be reformulated as
\begin{eqnarray} \label{eq36}
\overline{I}_1\! =\! \mathbb{E}_{\bm\omega}\!\left\{\!\sum_{m=1}^M \!\left[
\phi_{(1)}
\left(\theta_m^{(1)}\!(\bm\omega);x_m\right)\!+\!{\phi}_{(2)}
\left(\theta_m^{(2)}\!(\bm\omega);y_m\right)\!\right]\!\right\}\!.\!
\end{eqnarray}

After some simple manipulations, the problem $(\sf P)$ can be reformulated as a convex optimization problem, i.e.,
\begin{eqnarray} \label{eq35}
\!\!\!\!\!\!\!\!\!\!\!\!\!\!\!\!\!\!\!\!\!\!\!\!\!\!\!\!\!\!\!\!\!\!\!\!\!\!\!\!\!\!\min_{\substack{P_{s,n}^{(1)}(\bm\omega),P_{s,n}^{(2)}(\bm\omega),P_{r,n}(\bm\omega), \\\theta_m^{(1)}(\bm\omega),\theta_m^{(2)}(\bm\omega),n\in\mathcal {N},~m\in\mathcal {M}}}~
\overline{I}_1
\end{eqnarray}
\vspace{-0.5cm}
\begin{eqnarray}
\text{s.t.}~&&\!\!\!\!\!\!\!\!\!\overline{R}_{1}\geq {R}_{\min},~~\overline{R}_{2}\geq {R}_{\min}\\
&&\!\!\!\!\!\!\!\!\!\mathbb{E}_{\bm\omega}\left\{\sum_{n = 1}^N \left[\overline{P}_{s,n}^{(1)}(\bm\omega)+ P_{s,n}^{(2)}(\bm\omega)\right]\right\}\!\leq\! {P}^s_{\max} \\
&&\!\!\!\!\!\!\!\!\!\mathbb{E}_{\bm\omega}\left\{\sum_{n = 1}^N P_{r,n}(\bm\omega)\right\}\leq {P}^r_{\max}\\
%\end{align}
%\end{subequations}
%\setcounter{equation}{47(d)}
%\begin{subequations}
%\begin{align}
&&\!\!\!\!\!\!\!\!\! P_{s,n}^{(1)}(\bm\omega),P_{s,n}^{(2)}(\bm\omega),P_{r,n}(\bm\omega)\geq 0,~n\in\mathcal {N}\label{eq45}\\
&&\!\!\!\!\!\!\!\!\! 0\leq\theta_m^{(1)}(\bm\omega)
\leq\alpha,0\leq\theta_m^{(2)}(\bm\omega)\leq 1-\alpha,~m\in\mathcal {M},~~\label{eq46}
\end{eqnarray}
where $\overline{R}_{1},\overline{R}_{2}$ are determined by
\begin{eqnarray}\label{eq72}
\!\!\!\!\!\!\!\!\!\!&&\overline{R}_{1}\!=\!W\!\!\sum_{m\in\mathcal {M}}\sum_{n\in\mathcal {N}_m}\!\!\mathbb{E}_{\bm\omega}\!\!\left[\!{\theta_m^{(2)}}(\bm\omega)
\log_2\!\left(\!1\!+\!\frac{
P_{s,n}^{(2)}(\bm\omega)g_n^{s,d}}{\theta_m^{(2)}(\bm\omega)}\right)\right.\nonumber\\
\!\!\!\!\!\!\!\!\!\!\!&&~~~~~\left.+{\theta_m^{(1)}}(\bm\omega)
\log_2\!\!\left(\!1\!+\!\frac{
P_{s,n}^{(1)}(\bm\omega)\max\{g_n^{s,r},g_n^{s,d}\}}{\theta_m^{(1)}(\bm\omega)}\!\right)\!\right]\!,\!\!\\
\!\!\!\!\!\!\!\!\!\!\!&&\overline{R}_{2}\!=\!W\!\!\sum_{m\in\mathcal {M}}\sum_{n\in\mathcal {N}_m}\!\!\mathbb{E}_{\bm\omega}\!\!\left[\!
{\theta_m^{(1)}(\bm\omega)} \log_2\!\left(\!1\!+\!\frac{
P_{s,n}^{(1)}(\bm\omega)g_n^{s,d}}{\theta_m^{(1)}(\bm\omega)}\!\right)
\right.\nonumber\\
\!\!\!\!\!\!\!\!\!\!&&~~~~~\left.+
\theta_m^{(2)}(\bm\omega)
\log_2\!\left(\!1\!+\!\frac{P_{s,n}^{(2)}(\bm\omega)g_n^{s,d}\!+\!
P_{r,n}(\bm\omega)g_n^{r,d}}{\theta_m^{(2)}(\bm\omega)}\!\right)\!\!\right]\!.\!\!\label{eq73}
\end{eqnarray}

\subsection{The optimal solution of $(\sf P)$}\label{sec3}
By solving the KKT conditions of the derived convex optimization problem \eqref{eq35}-\eqref{eq46}, we derived the optimal solution for each realization of the NSI $\bm \omega$ and fixed dual variables \cite{YinSunTSP11}:

The optimal value of the ratio ${P_{s,n}^{(1)}(\bm \omega)}/{\theta_m^{(1)}(\bm \omega)}$ is given by
\begin{eqnarray}\label{eq60}
\!\frac{P_{s,n}^{(1)}(\bm \omega)}{\theta_m^{(1)}(\bm \omega)}\!=\!\textrm{positive
root $x$ of (\ref{eq59}) if it exists, otherwise }0,\!\!
\end{eqnarray}
and the root $x$ is determined by
\begin{eqnarray} \label{eq59}
\frac{\zeta \max\{g_n^{s,r},g_n^{s,d}\}}{1+x\max\{g_n^{s,r},g_n^{s,d}\}}+\frac{\sigma
g_n^{s,d}}{1+xg_n^{s,d}}=\varepsilon\ln2,
\end{eqnarray}
which is equivalent with a quadratic equation with closed-form
solutions.

The optimal values of the ratios ${P_{s,n}^{(2)}(\bm \omega)}/{\theta_m^{(2)}(\bm \omega)}$ and
${P_{r,n}(\bm \omega)}/{\theta_m^{(2)}(\bm \omega)}$ are given by
\begin{eqnarray}\label{eq67}
&&\frac{P_{s,n}^{(2)}(\bm \omega)}{\theta_m^{(2)}(\bm \omega)}=\left(\frac{\zeta}{(\varepsilon-\eta
g_n^{s,d}/g_n^{r,d})\ln2}-\frac{1}{g_n^{s,d}}\right)^+,\\
&&\frac{P_{r,n}(\bm \omega)}{\theta_m^{(2)}(\bm \omega)}=\frac{\sigma}{\eta\ln2
}-\frac{1}{g_n^{r,d}}-\frac{P_{s,n}^{(2)}(\bm \omega)g_n^{s,d}}{\theta_m^{(2)}(\bm \omega)g_n^{r,d}},\label{eq68}
\end{eqnarray}
with $(\cdot)^+ \triangleq \max(\cdot,0)$, if
$P_{r,n}(\bm \omega)>0$ is
satisfied. Otherwise, if $P_{r,n}(\bm \omega)=0$, we obtain
\begin{eqnarray}\label{eq65}
&&\frac{P_{s,n}^{(2)}(\bm \omega)}{\theta_m^{(2)}(\bm \omega)}=\left(\frac{\zeta+\sigma}{\varepsilon\ln2}-\frac{1}{g_n^{s,d}}\right)^+,\\
&&\frac{P_{r,n}(\bm \omega)}{\theta_m^{(2)}(\bm \omega)}=0.\label{eq75}
\end{eqnarray}

The optimal value of $\theta_m^{(1)}(\bm \omega)$ is determined as
\begin{eqnarray} \label{eq12}
\!\!\!\!\!\!\!\!\!\left\{\begin{array}{l}\!\!\theta_m^{(1)}(\bm \omega)
\!=\!\left[-\frac{1}{(\lambda+\mu)T_f}\!\ln\!\left\{1\!-\!\frac{\lambda+\mu}{\lambda}\sum\limits_{n\in
\mathcal {N}_m}\left[{\sigma}
f\left(g_n^{s,d}\frac{P_{s,n}^{(1)}(\bm \omega)}{\theta_m^{(1)}(\bm \omega)}\right)\right.\right.\right.\\
~~\!\left.\left.\left.+{\zeta}
\!f\left(\max\{g_n^{s,r},g_n^{s,d}\}\frac{P_{s,n}^{(1)}(\bm \omega)}{\theta_m^{(1)}(\bm \omega)}\!\right)\right]\right\}\right]^{\alpha}_0,
\textrm{if}~ x_m = 0,\\
\!\!\theta_m^{(1)}(\bm \omega)
\!=\!\left[\alpha\!+\!\frac{1}{(\lambda+\mu)T_f}\!\ln\!\left\{\frac{\lambda+\mu}{\mu}\!
\!\sum\limits_{n\in \mathcal {N}_m}\left[{\sigma}
f\!\left(g_n^{s,d}\frac{P_{s,n}^{(1)}(\bm \omega)}{\theta_m^{(1)}(\bm \omega)}\right)\right.\right.\right.\\
~~~~~\!\left.\left.\left.+{\zeta}f\left(\max\{g_n^{s,r},g_n^{s,d}\}\frac{P_{s,n}^{(1)}(\bm \omega)}{\theta_m^{(1)}(\bm \omega)}\right]\!-\!\frac{\lambda}{\mu}\right\}\!\right)\right]^{\alpha}_0,\textrm{if}~
x_m =
1,\end{array}\right.\!\!\!\!\!\!\!\!\!\!\!\!\!\!\!\!\!\!\!\!\!\!\!\!
\end{eqnarray}
where the value of ${P_{s,n}^{(1)}(\bm \omega)}/{\theta_m^{(1)}(\bm \omega)}$ is given by (\ref{eq60}), $f(x)\triangleq\log_2\left(1+x\right)-\frac{x}{(1+x)\ln2}$, $[x]^{y}_0\triangleq\min\{\max\{x,0\},y\}$, and
$\ln(x)$ is extended to take the value $-\infty$ for
$x\in(-\infty,0]$ to simplify the formulations. The optimal value of
$\theta_m^{(2)}(\bm \omega)$ is given by
\begin{eqnarray} \label{eq13}
\!\!\!\!\!\!\!\!\!\!\!\!\left\{\begin{array}{l} \!\!\theta_m^{(2)}(\bm \omega)
\!=\!\left[-\frac{1}{(\lambda+\mu)T_f}\!\ln\!\left\{1\!-\!\frac{\lambda\!+\!\mu}{\lambda}\!
\!\sum\limits_{n\in \mathcal
{N}_m}\left[\zeta f\left(g_n^{s,d}\frac{P_{s,n}^{(2)}(\bm \omega)}{\theta_m^{(2)}(\bm \omega)}\right)\right.\right.\right.\\
~\left.\left.\left.+{\sigma}
f\left(\!g_n^{s,d}\frac{P_{s,n}^{(2)}(\bm \omega)}{\theta_m^{(2)}(\bm \omega)}\!+g_n^{r,d}\frac{P_{r,n}(\bm \omega)}{\theta_m^{(2)}(\bm \omega)}\right)\right]\right\}\right]^{1-\alpha}_0\!\!\!\!,\textrm{if}~ y_m = 0,\\
\!\!\theta_m^{(2)}(\bm \omega)
\!=\!\left[1-\alpha\!+\!\frac{1}{(\lambda\!+\!\mu)T_f}\!\ln\!\left\{\frac{\lambda\!+\!\mu}{\mu}
\!\sum\limits_{n\in \mathcal
{N}_m}\!\!\left[\zeta f\left(g_n^{s,d}\frac{P_{s,n}^{(2)}(\bm \omega)}{\theta_m^{(2)}(\bm \omega)}\right)\right.\right.\right.\\
\!\left.\left.\left.+{\sigma}f\left(\!g_n^{s,d}\frac{P_{s,n}^{(2)}(\bm \omega)}{\theta_m^{(2)}(\bm \omega)}\!+g_n^{r,d}\frac{P_{r,n}(\bm \omega)}{\theta_m^{(2)}(\bm \omega)}\right)\right]\!-\!\frac{\lambda}{\mu}\right\}\right]^{1-\alpha}_0,\textrm{if}~
y_m =
1,\end{array}\right.\!\!\!\!\!\!\!\!\!\!\!\!\!\!\!\!\!\!\!\!\!\!\!\!\!\!\!
\end{eqnarray}
where the values of ${P_{s,n}^{(2)}(\bm \omega)}/{\theta_m^{(2)}(\bm \omega)}$ and
${P_{r,n}(\bm \omega)}/{\theta_m^{(2)}(\bm \omega)}$ are given by
(\ref{eq67})-(\ref{eq75}). Substituting (\ref{eq12})-(\ref{eq13})
into (\ref{eq60})-(\ref{eq75}), the optimal values of
$P_{s,n}^{(1)}(\bm \omega),P_{s,n}^{(2)}(\bm \omega), P_{r,n}(\bm \omega)$ are derived.

We now optimize the dual variables $\bm\nu\triangleq\{\zeta,\sigma,\varepsilon,\eta\}^T$ by the subgradient method \cite{YinSunTSP11}, where the subgradient ${\bm h}(\bm\nu)$ at the dual point $\bm\nu$ is given by
\begin{eqnarray}\label{eq42}
{\bm h}(\bm\nu)\!=\!\left[\!\begin{array}{l}({R}_{\min}-\overline{R}_{1}^\star)/W\\({R}_{\min}-\overline{R}_{2}^\star)/W\\
\sum_{n=1}^N
\mathbb{E}_{\bm\omega}\left\{P_{s,n}^{(1)\star}(\bm\omega)+P_{s,n}^{(2)\star}(\bm\omega)\right\}-{P}_{\max}^s\\\sum_{n=1}^N
\mathbb{E}_{\bm\omega}\left\{P_{r,n}^\star(\bm\omega)\right\} -{P}_{\max}^r\end{array}\!\right]\!,\!
\end{eqnarray}
where $P_{s,n}^{(1)\star}(\bm\omega)$, $P_{s,n}^{(2)\star}(\bm\omega)$ and $P_{r,n}^\star(\bm\omega)$ are derived throught \eqref{eq60}-\eqref{eq13} at the dual point $\bm\nu$, and $\overline{R}_{1}^\star$ and $\overline{R}_{2}^\star$ are the corresponding rate values in \eqref{eq72} and \eqref{eq73}, respectively.

%\begin{table}
%\caption{The dynamic transmission strategy DTS-1.} \label{tab3}
%\centering
%\begin{tabular}{||l||}
%\hline 1 The relay network estimates the channel fading distribution and \\
%~~~$\lambda$, $\mu$ in idle frames; \\
%2 the destination solves the dual problem off-line based on \cite[Tab. 1]{Yin2};\\
%3 in the $k$th frame,\\
%~~~3.1 the destination computes the power densities (\ref{eq60})-(\ref{eq75}), and the \\
%~~~~~~~~intermediate results (\ref{eq101})-(\ref{eq102});\\
%~~~3.2 the destination sends these results to the source and relay ahead\\
%~~~~~~~~of spectrum sensing;\\
%~~~3.3 the source and relay compute the transmission time (\ref{eq12})-(\ref{eq13})\\
%~~~~~~~~based on (\ref{eq116})-(\ref{eq102}) and the sensing outcomes $x_m^{(1)},x_m^{(2)}$;\\
%~~~3.4 the source and relay queue their data and encode messages\\
%~~~~~~~~according to \cite[Appendix B]{Yin2}; \\
%~~~3.5 the destination acquires $x_m^{(1)},x_m^{(2)}$ and then decode the messages.\\
%\hline
%\end{tabular}
%\end{table}
\subsection{Real-time implementations}\label{sec9}
%The above dual optimization solution enables an efficient dynamic
%transmission strategy with reduced computation burden at the source
%and a short sensing-transmission delay.
In the following, we show that dual variable $\bm\nu$ can be optimized off-line, which reduces the amount of real-time computations greatly. Moreover, by utilizing the structure of the optimal solution \eqref{eq60}-\eqref{eq13}, the primal solutions $\{P_{s,n}^{(1)}(\bm\omega),P_{s,n}^{(2)}(\bm\omega),P_{r,n}(\bm\omega),\theta_m^{(1)}(\bm\omega),
\theta_m^{(2)}(\bm\omega)\}$ can be updated on-line efficiently based on real-time NSI $\bm\omega$ of each frame, while generating quite short sensing-transmission delay.
\subsubsection{Off-line dual optimization}
These expectations \eqref{eq72}, \eqref{eq73} and \eqref{eq42}
do not have closed-form expressions. In practice, one can
compute the subgradient ${\bm h}(\bm\nu)$ by means of Monte Carlo
simulations. Specifically, one may randomly generate a set of
realizations of the NSI $\bm\omega$ following the distributions of the
CQIs and sensing outcomes. Then, the expectation terms in
\eqref{eq72}, \eqref{eq73} and \eqref{eq42} can be obtained by computing \eqref{eq60}-\eqref{eq13}, \eqref{eq72} and \eqref{eq73} for each realization of $\bm\omega$, and then
averaging the corresponding terms in \eqref{eq72}, \eqref{eq73} and \eqref{eq42} over these realizations. By this, the subgradient updates with high computation burden
can be performed off-line without using real-time NSI. %After
%we solve the dual problem, the optimal dual solution is used for
%calculating the optimal primal solution based on the instant NSI.
%We first consider the dual optimization based on subgradient method.
%When
%the distribution of NSI changes due to user movement or variations
%of ad-hoc traffic, previously obtained dual optimal solution can be
%used as the initial point of the current dual iterations to
%accelerate the convergence speed.
\subsubsection{On-line primal solution update}
In practice, the
BS (destination) acquires the CQI
$\{g_{n}^{s,r}(l),g_{n}^{s,d}(l),g_{n}^{r,d}(l)\}_{n=1}^N$ of Frame $l$ even before Frame $l$ starts through
prediction \cite{ZhangYan_prediction}, if the wireless channel varies slowly across the frames. Therefore, the BS can compute the ratio
$\frac{P_{s,n}^{(1)}(\bm\omega_l)}{\theta_m^{(1)}(\bm\omega_l)}$,
$\frac{P_{s,n}^{(2)}(\bm\omega_l)}{\theta_m^{(2)}(\bm\omega_l)}$ and
$\frac{P_{r,n}(\bm\omega_l)}{\theta_m^{(2)}(\bm\omega_l)}$ according
to \eqref{eq59}-\eqref{eq75} in Frame $l-1$. While the sensing
outcome $x_m(l)$ and $y_m(l)$ is still unknown at the BS at this moment,
the BS can compute $\theta_m^{(1)}(\bm\omega_l)$ and $\theta_m^{(2)}(\bm\omega_l)$ in
\eqref{eq12} and \eqref{eq13} by considering the two possible values of
of $x_m(l)$ and $y_m(l)$, respectively. Then, the
BS sends
$\frac{P_{s,n}^{(1)}(\bm\omega_l)}{\theta_m^{(1)}(\bm\omega_l)}$,
$\frac{P_{s,n}^{(2)}(\bm\omega_l)}{\theta_m^{(2)}(\bm\omega_l)}$ and
the possible vales of $\theta_m^{(1)}(\bm\omega_l)$ and
$\theta_m^{(2)}(\bm\omega_l)$ to the MT before Frame $l$
starts, and sends
$\frac{P_{r,n}(\bm\omega_l)}{\theta_m^{(2)}(\bm\omega_l)}$ and the
possible values of $\theta_m^{(2)}(\bm\omega_l)$ to the relay
before Phase 2 of Frame $l$ starts.

After receiving the feedbacks from the destination, the MT performs spectrum sensing at the beginning of Phase 1, and then selects the value of
$\theta_m^{(1)}(\bm\omega_l)$ according to the sensing outcome
$x_m(l)$. After Phase 1 of Frame $l$, the MT and relay node
perform spectrum sensing again at the beginning of Phase 2, and then
selects the value of $\theta_m^{(2)}(\bm\omega_l)$ in
accordance with the sensing outcomes $y_m(l)$. Therefore
the MT and relay nodes can transmit information signals right
after spectrum sensing with almost no sensing-transmission delay.

\begin{figure}[!ht]
    \centering
        \includegraphics[width=3.5in]{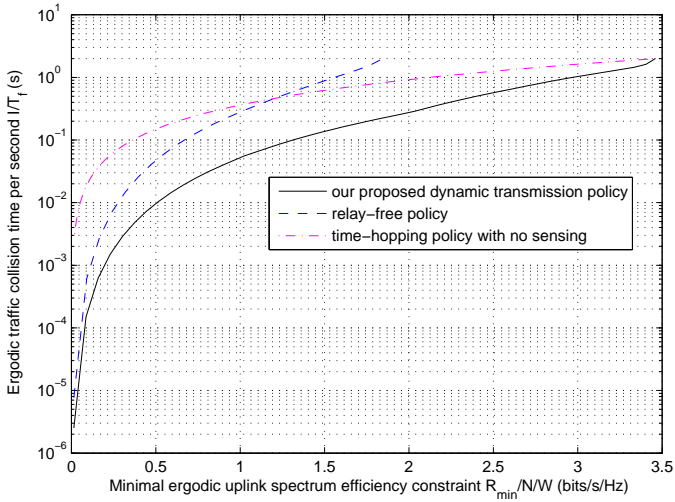}
        \caption{The interference mitigation performance of different spectrum sharing policies.}
        \label{fig6}
\end{figure}

\section{Numerical experiments}\label{sec1}
%\subsection{Reference policies}
We now compare our dynamic transmission policy with 2 reference
policies:

1. The first one is a \emph{relay-free policy} \cite{Geirhofer_Mobile_computing}, where the source transmits signals directly to the destination without using the relay node.

2. We then consider a \emph{time-hopping} random access policy with no spectrum sensing \cite{Time_hoppingTWC09}, where the CRN's transmission time is chosen randomly in each frame like frequency hopping. In this policy, the transmission time of the CRN satisfies $\theta_n^{(i)}(\bm\omega_l)=\theta$ for $n=1,\ldots, N$, $i=1,2$ and all frame index $l$, and the transmission powers of the source and relay are allocated optimally to maximize $\overline{R}_{DF}$.

%3. Finally, we consider a \emph{probabilistic access strategy},
%where the CRN accesses the whole transmission period with certain
%probability. More specifically, if $x_m^{(1)} =0$ ($x_m^{(2)}=0$), the CRN
%accesses the sub-channels $n\in\mathcal {N}_m$ in the BC (MAC) period
%with probability $p_0$. If $p_0=1$, the CRN starts to transmit in
%the sub-channels with busy sensing result, i.e. $x_m^{(1)} =1$
%($x_m^{(2)}=1$), by probability $p_1$. If $p_0<1$, we let $p_1=0$. The
%source and relay power is allocated to maximize $R_{DF,average}$.
%The value of $p_0$ and $p_1$ are chosen to satisfies
%$R_{DF,average}=R_{\min}$.

%\begin{table} \caption{Simulation parameters}
%\label{tab5} \centering
%\begin{tabular}{||c|c||c|c||}
%\hline Parameters & Value & Parameters & Value\\
%\hline $N$ & 16 & $\lambda T$ & 1 \\
%\hline $M$ & 4 & $\nu T$ & 1\\
%\hline $\alpha$ & 0.5 & & \\
%\hline
%\end{tabular}
%\end{table}
%\begin{figure}[!ht]
%    \centering
%        \includegraphics[width=2.5in]{2.eps}
%        \caption{The locations
%of the source, relay and destination.}
%        \label{fig3}
%\end{figure}

%\subsection{The numerical results}
We consider that the source, relay and destination stands in a line and the relay locates in the middle of the source and destination. The CRN has $N=16$ sub-channels, and the ad-hoc network has $M=4$ bands. Thus, each ad-hoc band overlaps with 4 CRN sub-channels. The channel gain between every two nodes of the CRN at each sub-channel can be decomposed into a small-scale Rayleigh fading and a large-scaled path loss component with a path-loss factor of 4. The small-scale fading are i.i.d. across the sub-channels to simulate a frequency-selective environment. We assume that the power constraints of the source and relay nodes are the same and the signal-to-interference-plus-noise ratio (SINR) of the source-destination link is $
\frac{P^s_{\max}\mathbb{E}\{g_n^{s,d}\}}{N}=5$dB. The parameters of ad-hoc traffic model satisfy $\mu T_f = \lambda T_f = 1$. The value of $\alpha$ is chosen to be $0.5$.

%We assume that the channel gain $g_n^{s,r},g_n^{r,d},g_n^{s,d}$ satisfies
%Rayleigh distribution. The average SINR of the S-D link is given by
%${P_{\max}^s\mathbb{E}\{g_n^{s,d}\}}/{N}=5$dB. The relay locates right in
%the middle of the source and destination, as shown in Fig.
%\ref{fig3}. We further assume that $P_{\max}^s=P_{\max}^r$ and the
%path-loss factor is 4. Since our strategy can achieve a quite short
%sensing-transmission time delay $\delta T$, the value of
%$\delta$ is approximately chosen as 0. Other system parameters
%are listed in Tab. \ref{tab5}.

Figure \ref{fig6} illustrates the interference mitigation performance of the spectrum sharing policies. We find
that our policy achieves better interference mitigation
performance than the reference policies. More specifically, the
relay-free policy is slightly worse than our policy in low
spectrum efficiency region. However, if the required uplink spectrum
efficiency is relative high, the interference mitigation performance
of relay-free policy is quite poor, because of its relative low
capacity. The spectrum efficiency of our policy is $80\%$ higher
than that of the relay-free policy, when the ergodic traffic collision
time per second is larger than $0.01$s. The time-hopping policy has quite
poor performance for relative low spectrum efficiency, because it has not utilized the spectrum sensing results.

%\subsection{The varying rate of ad-hoc traffic state}
%We define the varying rate of ad-hoc traffic state $\varsigma>0$ as
%the average number of ad-hoc transmission intervals in each frame,
%i.e.
%\begin{eqnarray}
%\varsigma=\frac{T}{\frac{1}{\lambda}+\frac{1}{\mu}}.
%\end{eqnarray}
%The case of $\varsigma\approx0$ means that the idle/busy state of
%the ad-hoc traffic hardly varies in a frame, while very large
%$\varsigma$ suggests that the ad-hoc traffic is almost
%unpredictable.
%
%
%The simulation results for different values of $\varsigma$ and
%$R_{\min}$ are illustrated in Fig \ref{fig7}-\ref{fig9} while
%maintaining $\lambda=\mu$. If $\varsigma\approx0$ and $R_{\min}$ is
%relative small, the sensing-based policies achieve better
%performance than the sensing-free policy. Moreover, our optimal
%policy generates much less interference than the relay-free policy
%and probabilistic access policy, because of the incremental data
%rate of DF relaying and the benefit of flexible resource allocation
%based on interference prediction.
%
%
%If $R_{\min}$ is close to the maximal data rate, these access
%policies need to utilize most of the transmission time to achieve
%the QoS requirement, thus the interference mitigation performance is
%relative poor. Since the capacity of direct S-D transmission is much
%lower than the achievable data rate of DF relaying, the relay-free
%policy becomes ineffective for $R_{\min}=1.2$ nat/s/Hz while the DF
%relaying based policies still work. If $\varsigma$ is relative
%large, all the access policies have poor performance because the
%ad-hoc traffic is almost unpredictable.

\section{Conclusions}\label{sec14}
This paper studied a spectrum sharing scenario between cooperative
relay and ad-hoc networks. A dynamic transmission policy of the CRN is proposed
which requires little real-time computation and guarantees high traffic prediction accuracy. The benefits of spectrum sensing and cooperative relay techniques are demonstrated by our numerical experiments. %We are currently considering
%generalizing this work to the cases of general ad-hoc traffic model
%and imperfect spectrum sensing.

%Real-time implementations for average resource allocation have been
%proposed. The computational burden of the source (mobile terminal) is
%quite small and the signaling delay and computation time has little
%impact on the accuracy of short term ad-hoc traffic prediction. The
%key idea to achieve this is making use of the analytical structure
%of the optimal solution. Cooperative relaying strategy is beneficial
%for both throughput enhancement and interference mitigation.
%
%
%
%Our simulation results show that flexible source allocation based on
%spectrum sensing and interference prediction achieves good
%performance, if the state of ad-hoc traffic varies slowly and the
%QoS requirement of the cooperative relay network is not too
%stringent. Otherwise, it is quite hard to suppress the interference.

\section*{Acknowledgement}
The authors would like to thank Yongle Wu, Ying Cui and Ness B. Shroff for constructive discussions about this work.
\bibliographystyle{IEEEtran}
\vspace{-0.2cm} \footnotesize
\bibliography{CR_relay_11}
\end{document}